\documentclass[a4paper,11pt]{article}
\usepackage{jcappub} 

\usepackage{lineno}
\usepackage{subcaption}
\usepackage{graphicx}
\usepackage[normalem]{ulem}
\usepackage{rotating} 

\usepackage{bm, color, colortbl}
\usepackage[citecolor=blue]{hyperref}
\definecolor{linkcolor}{rgb}{0.6,0,0}
\definecolor{citecolor}{rgb}{0,0,0.75}
\definecolor{urlcolor}{rgb}{0.12,0.46,0.7}

\hypersetup{colorlinks,linkcolor=linkcolor,citecolor=citecolor,urlcolor=urlcolor}  
\definecolor{kwrefcolor}{rgb}{0.6,0,0.6}

\newcommand{\wf}{WF}

\title{\boldmath Wavelet Flow For Extragalactic Foreground Simulations}

\author[a,b,c]{M. Mebratu,}
\author[b,c]{W.~L.~K. Wu}
\affiliation[a]{Stanford University Department of Physics,\\Stanford, CA, 94305, U.S.A}
\affiliation[b]{SLAC National Accelerator Laboratory,\\Menlo Park, CA, 94306, U.S.A}
\affiliation[c]{Kavli Institute for Particle Astrophysics and Cosmology,\\Stanford, CA, 94305, U.S.A}

\emailAdd{mmebrat1@stanford.edu}
\emailAdd{wlwu@stanford.edu}

\abstract{Extragalactic foregrounds in cosmic microwave background (CMB) observations are both a source of cosmological and astrophysical information and a nuisance to the CMB. 
Effective field-level modeling that captures their non-Gaussian statistical distributions is increasingly important for optimal information extraction, particularly given the low-noise observations from current and upcoming experiments.
We explore the use of Wavelet Flow~(\wf) models to tackle the novel task of modeling the field-level probability distributions of multi-component CMB secondaries and foregrounds.
Specifically, we jointly train correlated CMB lensing convergence ($\kappa$) and cosmic infrared background (CIB) maps with a
\wf\ model and obtain a network that statistically recovers the input to high accuracy---the trained network generates samples of $\kappa$ and CIB fields whose average power spectra are within a few percent of the inputs across all scales, and whose Minkowski functionals are similarly accurate compared to the inputs.
Leveraging the multiscale architecture of these models, we fine-tune both the model parameters and the priors at each scale independently, optimizing performance across different resolutions. 
These results demonstrate that \wf\ models can accurately simulate correlated components of CMB secondaries,
supporting improved analysis of cosmological data.
Our code and trained models can be found on this \href{https://github.com/matiwosm/HybridPriorWavletFlow.git}{GitHub repo}.}

\begin{document}
\maketitle
\flushbottom

\section{Introduction}
\label{sec:intro}

Observations of the cosmic microwave background (CMB) set the foundation of the standard model of cosmology~\cite{plank2013, plank2015, plank2018} and are pivotal in understanding the formation of cosmic structures~\cite{cmblensingreview, plank2013lensing, planks2018lensing, spt-sz-sigamtsz, plank2015sigamtsz, plank2013cib}. As CMB observations become more precise, it is increasingly important to accurately model the various sources of foreground emissions in the mm-wave wavelengths~\cite{cibtszradio, vanengelenfgbias, radiocibimpactontszandlensing}.

The main extragalactic foreground emissions are the cosmic infrared background (CIB), the thermal and kinetic Sunyaev-Zeldovich effects (tSZ/kSZ), and radio sources~\cite{tszoriginal, cibdetection, cibreview, dezotti}. These foregrounds present both challenges and opportunities for cosmological analyses. 
Proper characterization allows us to remove their biases to CMB measurements~\cite{vanengelenfgbias, actdr6lensing} and harness them as sources of information for studies of cosmology and structure formation~\cite[e.g.][]{plankGlensingxCIB, hillpajer_tsz, baryons_actsdss, yxgal_sanchez}.

Because foreground probability distributions are analytically intractable, their inference requires simulation-based inference~(SBI) methods to extract the complete or near-complete information content from the foregrounds. However, generating realistic foreground simulations presents significant computational challenges as these components exhibit complex non-Gaussian statistics, spatial correlations across different scales, and interdependencies between different foreground components~\cite{websky, yuuki, halfdome}. Traditional simulation approaches often involve computationally expensive hydrodynamical simulations or N-body simulations followed by semi-analytic modeling of baryons, making them impractical for many inference tasks that require numerous realizations~\cite{deeplearningfg}.

In this work, we address the simulation generation challenge using Normalizing Flows~\cite{NflowDinh2014NICENI, Nflowreview, NflowDinh2016DensityEU, NflowJimenezRezende2015VariationalIW, NflowKingma2018GlowGF, Nflowadam, Nflowternf, NflowGalaxy–Halo, NflowH1flow, NflowStachurski2023CosmologicalIU}, a class of deep generative models that can efficiently learn complex probability distributions by transforming simple base distributions into complex ones with a learned bijective function, allowing both the generation of new realizations and the density estimation of the target distribution.

Specifically, we employ Wavelet Flow (\wf)~\cite{waveletflowCS, waveletflowCosmology}, a specialized implementation of Normalizing Flows. The key innovation of \wf\ is its use of the discrete wavelet transform (DWT) to decompose images into high- and low-frequency components, allowing separate flow transformations to be applied at each scale. This multiscale approach facilitates dimensional reduction for more efficient training and enables the model to capture both large-scale correlations and small-scale features present in cosmological fields, while allowing for the fine-tuning of deep learning models independently at each scale during training.

We apply \wf\ on simulated maps of CMB foregrounds. 
The foreground fields are correlated with each other, as well as with the CMB lensing convergence field $\kappa$, because they all trace the same underlying large-scale structure.
Previous works~\cite{Nflowadam, waveletflowCosmology} have trained Normalizing Flows on $\kappa$ but are limited to a single component.
For analysis of real CMB observations that contain both the CMB and various foregrounds, it is essential to jointly model multiple components.
This work demonstrates its feasibility with a flow-based model for the first time.\footnote{Other kinds of generative models such as GANs have been applied to multiple components in a similar context~\cite{mmDL} but they do not learn explicit probability distributions. Although the focus of this work is on sampling, we choose Normalizing Flows because they enable the explicit calculation of the log-likelihood, which is essential for future density estimation and inference tasks.}
As a proof of concept, we apply WF to jointly model the lensing convergence $\kappa$ and the CIB, capturing their statistical properties and cross-correlations.

Modified priors have been shown to enhance the expressivity of Normalizing Flows~\cite{Nflowadam, ModpriorStimper2021ResamplingBD, ModpriorCrenshaw2024ProbabilisticFM, ModpriorPapamakarios2017MaskedAF}. We leverage the modular nature of \wf s to explore this potential by implementing different prior distributions across various scales. Our results indicate that this scale-dependent approach to prior selection can enhance model performance, offering a flexible framework that capitalizes on the multiresolution structure inherent to \wf\ models.

This work offers a promising pathway toward fast, accurate generation of multi-component CMB foreground simulations. 
The trained flows can be used as part of the simulator in simulation-based/likelihood-free inference, for generating covariance matrices that include beyond-Gaussian information, and as mock skies to test pipelines for CMB and large-scale structure analyses.
As a concrete example, for simulation-based inference as done in~\citep{muse3g}, the generated correlated $\kappa$-CIB maps can be input to the sky simulation step to make foreground-added lensed CMB  maps as part of the needed realistic simulation model. The learned, correlated probability distributions between the $\kappa$ and the CIB fields can be used as priors in the posterior model in the same framework.

This paper is organized as follows: In Section~\ref{sec:kappacib}, we provide more background on the fields we emulate with the flow-based model. In Section~\ref{sec:methods} we introduce the models we use in this work. In Section~\ref{sec:imple}, we discuss the implementation details, including data preprocessing and the specific architecture of our \wf\ models. Section~\ref{sec:Results} presents our findings, demonstrating the performance of our models in capturing the joint $\kappa$-CIB statistics across various scales. We analyze the summary statistics of our generated samples, comparing them to the input data. Finally, in Section~\ref{sec:DandC} we summarize our findings and discuss their potential for upcoming CMB and large-scale structure experiments. 

\section{CMB lensing convergence and the cosmic infrared background}\label{sec:kappacib}

The CMB, last scattered at redshift $z \approx 1100$, encodes information about primordial density fluctuations that seeded the growth of cosmic structures observed today~\cite{osti_2349447, plank2015}. 
As CMB photons traverse the universe, their paths are deflected by the intervening large-scale structure~(gravitational lensing) and they can be scattered by electrons~(Sunyaev-Zel'dovich effects).
Both processes encode additional cosmological information of the late universe~\cite{plank2013lensing, 2017tszcosmologyconstraints, spt-sz-sigamtsz, plankISW}. 
Furthermore, besides the lensed CMB and its secondaries, mm-wave observations contain emissions from galaxies and sources of different spectral types: dusty sources with positive spectral indices and synchrotron sources with negative spectral indices~\cite{sptsources, actsources}.
The dusty sources form the CIB, while the synchrotron sources that are typically bright in the radio wavelengths are called radio sources.
In this work, we focus specifically on the CMB lensing convergence $\kappa$ and the cosmic infrared background.

Gravitational lensing of the CMB arises as photons propagate through the inhomogeneous matter distribution. The effect is characterized by the lensing potential $\phi$, a two-dimensional projection of the 3D gravitational potential $\Phi$ along the line of sight. Mathematically, assuming flat geometry, $\phi$ is expressed as a weighted integral:

$$\phi(\hat{\mathbf{n}}) = -2 \int_0^{\chi_*} \! d\chi \, \frac{\chi_* - \chi}{\chi_* \chi} \, \Phi(\chi\hat{\mathbf{n}}, \chi),$$
where $\chi$ is the comoving distance and $\chi_*$ is the distance to the last scattering surface~\cite{cmblensingreview, Hu:2001kj}. The convergence field $\kappa$, derived from $\kappa = -\nabla^2 \phi / 2$, corresponds to a weighted integral of the matter overdensity $\delta$ and quantifies the projected mass distribution along the line of sight~\cite{cmblensingreview, lensingsim, Bartelmann:1999yn}. Gravitational lensing introduces statistical anisotropy (e.g., mode couplings) in the CMB temperature and polarization anisotropies, which are used to reconstruct the lensing field and its power spectrum~\cite{Hu:2001kj, planks2018lensing}.

The lensing convergence field can be effectively modeled as Gaussian for a broad range of scales that are relevant for lensing the CMB. At small scales ($\ell \gtrsim 2000$), non-linearities from gravitational collapse start to become important, which introduces non-Gaussianities to $\kappa$. 
As noise continues to decrease for CMB experiments, the small scales will become better measured. 
As such, accurate models of the probability distribution of the $\kappa$ field will enable more optimal analysis of data~\cite[e.g.][]{diffdelensing}.

The CIB represents the aggregated infrared light emitted by dust-obscured star-forming galaxies~\cite{cibreview, CASEY201445}. This diffuse background radiation peaks at far-infrared wavelengths and originates primarily from thermal dust emission in galaxies during the epoch of peak cosmic star formation~\cite{CASEY201445, plank2013cib}. Dust grains in these galaxies absorb ultraviolet and optical light from young stars and re-emit this energy at longer wavelengths, creating a signal that traces the dust-obscured star formation history~\cite{Maniyar2021, CASEY201445, Lagache:2005sw, plank2013cib}.

Of the foreground fields, the CIB is the most correlated with $\kappa$. 
This correlation comes from two factors: first, the star-forming galaxies, which contribute to the CIB, are predominantly located in massive dark matter halos that also impact the lensing potential~\cite{plankGlensingxCIB, Song_2003}; second, there is substantial overlap in the redshift ranges where the lensing kernel peaks and star formation is most active ($z \sim 1 - 3$)~\cite{plankGlensingxCIB, CASEY201445}. As one of the main extragalactic contaminants in CMB observations at frequencies above 100~GHz, the CIB must be carefully modeled for both analysis of the primary CMB and for CMB lensing~\cite{plank2018, yuuki, actdr6lensing, spt2018lensing}. 
In addition, the CIB (and the other foreground fields) are more non-Gaussian than CMB lensing and primary CMB fields~\cite{tszbispectrum, tszbispectrum_spt, cib_ng}, which make them one of the first non-Gaussian fields to model in CMB data analyses.

The growing need to have field-level descriptions of these fields' probability distributions for improved modeling in CMB data analyses, combined with the advances in neural network architectures and growing interests in field-level analysis of cosmological data motivate us to explore flow-based networks for this problem.

\section{Methods}
\label{sec:methods}
We introduce the class of neural network we use for this work. We first discuss Normalizing Flows, a type of generative model, followed by the specific Glow~\cite{NflowKingma2018GlowGF} architecture we implement. We then describe \wf, an application of Normalizing Flows on multiple separate scales. Finally, we discuss the choice of prior distributions from which the flow model starts, and the ability to apply different priors for each scale to enhance network expressivity.

\subsection{Normalizing Flows}
\label{sec:nf}
Normalizing Flows provide a framework for modeling  probability distributions over continuous variables. Suppose \( x \) is a \( D \)-dimensional real vector. Our goal is to model an accurate distribution for \( x \). The basic principle of flow-based modeling involves representing \( x \) through a transformation \( T \) applied to a real vector \( u \), drawn from a base distribution \( p_u(u) \) (also referred to as the \textit{prior} distribution throughout this paper):

\begin{equation}
x = T(u) \quad \text{where} \quad u \sim p_u(u).
\end{equation}
 \( T \) can be parameterized by multiple parameters \( \phi_{1}, \phi_{2}, \ldots, \phi_{N} \). By stacking a series of such transformations, we can generate more expressive models that capture the training data's probability distributions. In our case, they are the simulated maps of $\kappa$ and CIB.

An important characteristic of flow-based models is that the transformation 
\(T\) must be both invertible and differentiable. Invertibility and differentiability are necessary for training and density estimation. This requirement imposes some limits on the expressiveness of these models~\cite{ModpriorStimper2021ResamplingBD}. 

Given these properties, the density of \( x \) can be derived using the change of variables formula:
\begin{equation}
p_x(x) = p_u(u) |\det J_T(u)|^{-1} \quad \text{where} \quad u = T^{-1}(x).
\end{equation}
This relationship can also be expressed using the Jacobian of \( T^{-1} \):
\begin{equation}
p_x(x) = p_u(T^{-1}(x)) |\det J_{T^{-1}}(x)|.
\end{equation}
The Jacobian \( J_T(u) \), a \( D \times D \) matrix, includes all partial derivatives of \( T \) and is defined as:
\begin{equation}
J_T(u) = \begin{bmatrix}
\frac{\partial T_1}{\partial u_1} & \cdots & \frac{\partial T_1}{\partial u_D} \\
\vdots & \ddots & \vdots \\
\frac{\partial T_D}{\partial u_1} & \cdots & \frac{\partial T_D}{\partial u_D}
\end{bmatrix}.
\end{equation}
In practical implementations, the transformation \( T \) (or \( T^{-1} \)) is often constructed using neural networks, with \( p_u(u) \) typically modeled as a simple distribution such as a multivariate normal.

\subsection{Glow}
\label{sec:glow}
In this work, we implement the Glow architecture, a variant of Normalizing Flows for high-dimensional data. Glow uses affine coupling layers as its primary transformation function, a technique common across many flow-based models~\cite{NflowKingma2018GlowGF, Nflowreview}.

Affine coupling layers split the input vector into two parts: one part remains unchanged while the other is transformed with an affine transformation conditioned on the unchanged part. For an input vector \(z\) split into \(z_1\) and \(z_2\): 

\begin{equation}
\begin{aligned}
y_1 &= z_1, \\
y_2 &= z_2 \odot \exp(s(z_1)) + t(z_1),
\end{aligned}
\end{equation}
where \(s\) and \(t\) are neural networks computing scale and translation factors. This construction ensures differentiability and invertibility, with the inverse operation:

\begin{equation}
\begin{aligned}
z_1 &= y_1, \\
z_2 &= (y_2 - t(y_1)) \odot \exp(-s(y_1)).
\end{aligned}
\end{equation}
An advantage of this design is the triangular structure of its Jacobian, making the determinant computation tractable:

\begin{equation}
\det\left(\frac{\partial y}{\partial z}\right) = \prod_i \exp(s(z_1)_i).
\end{equation}

What distinguishes Glow from other flow-based models are three key elements: (1)~``actnorm'' layers that perform activation normalization similar to batch normalization but with data-dependent initialization and consistent behavior during training and inference; (2) invertible 1×1 convolutions that serve as learnable permutation operations; and (3) a multi-scale architecture for processing data at different resolutions.

The standard multi-scale structure in Glow employs a ``squeeze and split'' approach. At each scale transition, spatial dimensions are reduced by a factor of two while channel dimensions increase by a factor of four (``squeezing''). 
Half of these channels are then ``split'' off to the latent space, while the remainder continue through subsequent transformations (see e.g. Figure 2b of~\cite{NflowKingma2018GlowGF}). 
This process creates a hierarchy where coarse features are captured at early splits and fine details at later stages. 
Our implementation skips this step and relies on the wavelet decomposition to process each scale, as described in the following section.

\subsection{Wavelet Flow}
\begin{sidewaysfigure}[htbp]
    \centering
    \includegraphics[width=\textwidth]{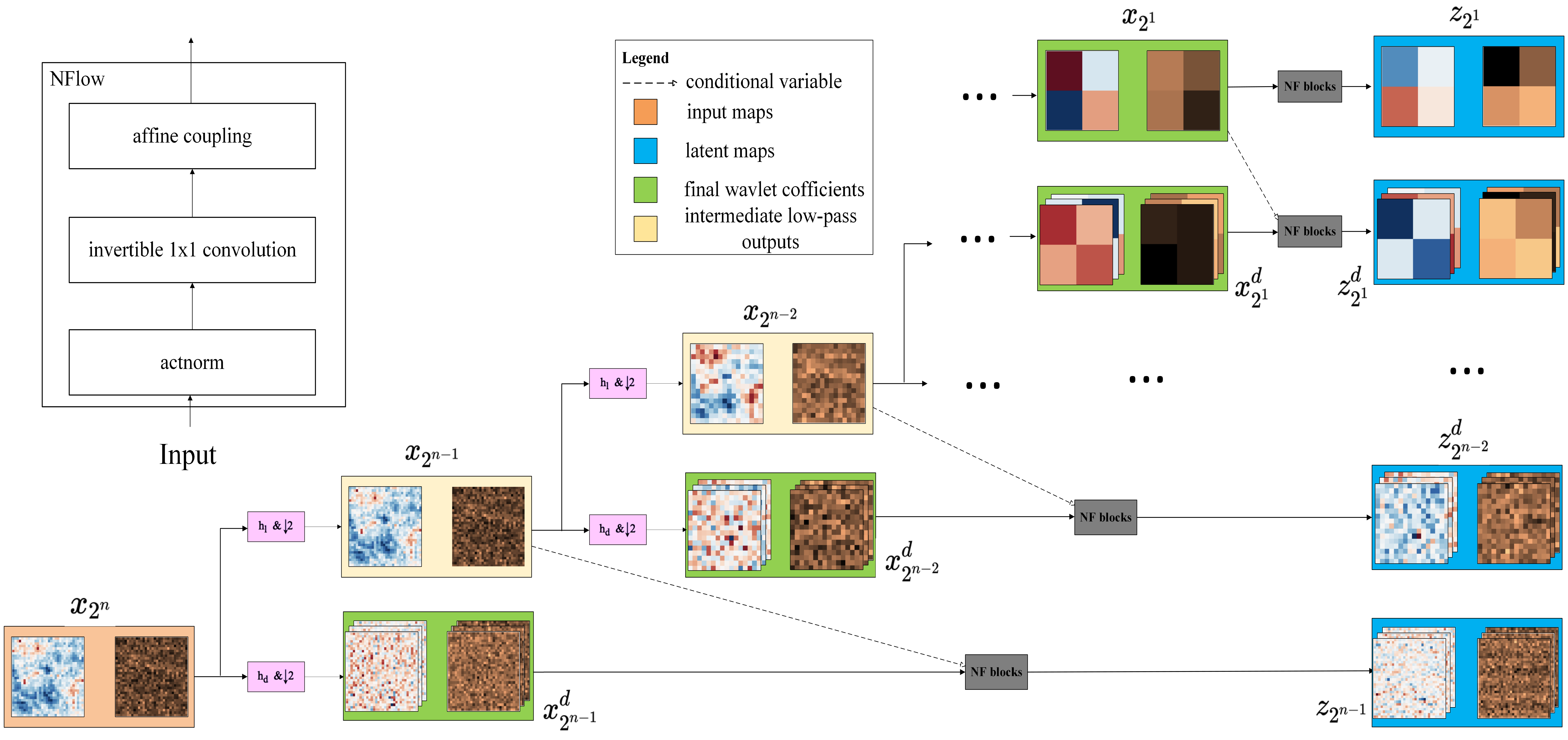}
    \caption{The diagram illustrates our implementation of the \wf\ model. Beginning with an input map \(x_{2^{n}}\), the map undergoes low-pass filtering \(h_l\), high-pass filtering \(h_d\), and a 2$\times$ downsampling. These steps are repeated at each iteration \(i\), where \(i\) ranges from 0 to \(n-1\). This processing results in a full wavelet representation of the map \{\(x^d_{2^{n-1}}, x^d_{2{^{n-2}}} , \dots, x^d_{2^1}, x_{2^1} \)\} (green boxes). These wavelet coefficients are then transformed through multiple Normalizing Flow (NF) blocks into Gaussian latent maps \{\(z^d_{2^{n-1}}, z^d_{2^{n-2}}, \dots, z^d_{2^1}, z_{2^1}\)\} (blue boxes). The original image can be reconstructed from its wavelet coefficients by applying the inverse wavelet transform \(x_{2^{n-i}} = h^{-1}(x_{2^{n-(i+1)}}, x_{2^{n-(i+1)}}^d)\) recursively until \(i = 0\). The two input maps are $\kappa$ (left) and CIB (right), with NF blocks being the Glow model from Section~\ref{sec:glow} without the multi-scale ``squeeze and split'' approach.}

    \label{fig:wavelet}
\end{sidewaysfigure}

\wf\ uses DWTs to decompose an image \( x_{2^n}~\in~\mathbb{R}^{2^n \times 2^n \times C} \) into multiple scales. This decomposition is done recursively, where at each level \( i \), the image is split into two components: low-frequency coefficients \( x_{2^{n-(i+1)}} \) obtained through a low-pass filter \( h_l(x_{2^{n-i}}) \), and high-frequency detail coefficients \( x_{2^{n-(i+1)}}^d \) obtained via a high-pass filter \( h_d(x_{2^{n-i}}) \). Here, \(i\) ranges from 0 to \(k\) where \(k\) is the base level, resulting in a final map with resolution \(2^{n-k}\). Throughout this work, following our implementation, we assume that \(k=n-1\), i.e., decomposition is performed down to the 2×2 pixel unless otherwise noted. For all wavelet transforms in this work, we employ the Haar wavelets~\cite{Haar1910}. The wavelet transform is invertible, meaning the original image can be reconstructed from the high-frequency and low-frequency components. The process is summarized in Figure~\ref{fig:wavelet}.

In \wf, we model the distribution of wavelet coefficients instead of the image directly. Applying a change of variables, the distribution \( p(x_{2^n}) \) of the image can be expressed as:

\begin{equation}
p(x_{2^n}) = p(\mathcal{H}(x_{2^n})) \left| \det \frac{\partial \mathcal{H}}{\partial x_{2^n}} \right|,
\end{equation}
where \( \mathcal{H} \) represents the wavelet transform. When using orthonormal wavelets, the determinant term simplifies to one, allowing us to factorize the distribution as:

\begin{equation}
p(x_{2^n}) = p(x_{2^1}) \prod_{i=1}^{n-1} p(x_{2^{n-i}}^d | x_{2^{n-i}}),
\label{eqn:prob_2n}
\end{equation}
This formulation allows us to model the distribution of the detail coefficients \( x_{2^{n-i}}^d \), conditioned on the lower resolution image \( x_{2^{n-i}} \). \wf\ is trained by maximizing the log-likelihood of the input data. Specifically, we maximize the $\log$ of Eqn.~\ref{eqn:prob_2n}.

An advantage of \wf s is that each conditional distribution of detail coefficients can be trained independently, which allows for parallelization and reduces memory requirements. As illustrated in Figure~\ref{fig:wavelet}, the conditional distributions of detail coefficients are modeled using a Glow-like model without the “squeeze and split” steps, which yields tractable log-probabilities for future density estimation.

Sampling from a \wf\ model begins by drawing latent variables \(z_{2^1}, z^d_{2^1}, \dots, z^d_{2^{n-2}}, z^d_{2^{n-1}}\) from their respective distributions \(p(z_{2^1}), p(z^d_{2^1}), \dots, p(z^d_{2^{n-2}}), p(z^d_{2^{n-1}})\). These latent variables are then mapped to wavelet coefficients \(x_{2^1}, x^d_{2^1}, \dots, x^d_{2^{n-2}}, x^d_{2^{n-1}}\) using the corresponding pre-trained Glow models. Specifically, for the low frequency component 
\[
x_{2^1}=T_k^{L1} \circ T_{k-1}^{L1} \circ \cdots \circ T_{1}^{L1}(z_{2^1}),
\]
and for the high-frequency components
\[x_{2^j}^d=T_k^{Hj} \circ T_{k-1}^{Hj} \circ \cdots \circ T_{1}^{Hj}(z_{2^j}^d | z_{2^j}),\]
where $T_k^{j}$ is the $k$-th NF block trained on the resolution level $j$ (corresponding to maps of size $2^j \times 2^j$) and the superscripts $L$ and $H$ indicate whether the model is trained on low-frequency or high-frequency wavelet coefficients. The original image is reconstructed recursively via inverse wavelet transforms, progressing from the lowest resolution to the highest.

\subsection{Hybrid Prior}
\label{sec:HCC}
Our \wf\ architecture uses Normalizing Flows to transform prior distributions into wavelet coefficient distributions across multiple scales. A key feature of this approach is its modularity; each decomposition level can be independently optimized with specialized priors or network architectures. This flexibility, combined with recent advances showing how modified priors can enhance flow expressivity~\cite{Nflowadam, ModpriorStimper2021ResamplingBD, ModpriorCrenshaw2024ProbabilisticFM, ModpriorPapamakarios2017MaskedAF}, makes \wf\ models promising for modeling complex distributions.

The prior distribution in a Normalizing Flow model refers to the base distribution from which the flow transformation begins. As introduced in Section~\ref{sec:nf}, this is the distribution $p_u(u)$ from which we sample before applying the transformation $T$ to generate our target distribution. In conventional implementations, this prior is typically a standard normal distribution, also called a ``White Noise'' (WN) prior, applied uniformly across all scales.

In this work, we explore ``Correlated'' priors that incorporate some aspects of the statistical structure present in our data. For a single component, these priors are constructed using a scale-dependent covariance:

\begin{equation}
\Sigma(\ell) = C(\ell), 
\end{equation}
where $C(\ell)$ represents the power spectrum of the component map at multipole $\ell$.
The Correlated prior distribution is defined as $\mathcal{N}(0, \Sigma(\ell))$ for each $\ell$ in the Fourier grid.

When training multiple components simultaneously, we extend this approach to a ``Component Correlated'' (CC) prior. For $N$ components, the covariance structure becomes:

\begin{equation}
\Sigma(\ell) = 
\begin{pmatrix}
C_{11}(\ell) & C_{12}(\ell) & \cdots & C_{1N}(\ell) \\
C_{21}(\ell) & C_{22}(\ell) & \cdots & C_{2N}(\ell) \\
\vdots & \vdots & \ddots & \vdots \\
C_{N1}(\ell) & C_{N2}(\ell) & \cdots & C_{NN}(\ell)
\end{pmatrix}
\end{equation}
where $C_{ij}(\ell)$ represents the power spectrum ($i=j$) or cross-spectrum ($i\neq j$) between components $i$ and $j$ at multipole $\ell$.

In wavelet space, we apply this framework to the wavelet coefficients. For $N$ components with wavelet decomposition, the covariance matrices use the power spectra of the wavelet coefficients rather than the original maps. For high-frequency detail coefficients, we maintain separate priors for each orientation $o$ (horizontal, vertical, and diagonal), assuming orthogonality between orientations:

\begin{equation}
\label{eq:cov_high_freq}
\Sigma_{\text{high}} = 
\begin{pmatrix}
\mathbf{C}_h & \mathbf{0} & \mathbf{0} \\
\mathbf{0} & \mathbf{C}_v & \mathbf{0} \\
\mathbf{0} & \mathbf{0} & \mathbf{C}_d
\end{pmatrix}
\end{equation}
where each $\mathbf{C}_o$ is an $N \times N$ matrix of auto- and cross-power spectra for orientation $o$:

\begin{equation}
\mathbf{C}_o = 
\begin{pmatrix}
C_{1o,1o} & C_{1o,2o} & \cdots & C_{1o,No} \\
C_{2o,1o} & C_{2o,2o} & \cdots & C_{2o,No} \\
\vdots & \vdots & \ddots & \vdots \\
C_{No,1o} & C_{No,2o} & \cdots & C_{No,No}
\end{pmatrix}.
\end{equation}
For low-frequency approximation coefficients, the same structure follows:

\begin{equation}
\label{eq:cov_low_freq}
\Sigma_{\text{low}} = 
\begin{pmatrix}
C_{1,\text{low},1,\text{low}} & \cdots & C_{1,\text{low},N,\text{low}} \\
\vdots & \ddots & \vdots \\
C_{N,\text{low},1,\text{low}} & \cdots & C_{N,\text{low},N,\text{low}}
\end{pmatrix}.
\end{equation}

The entries in the covariance matrices in Eqs.~\ref{eq:cov_high_freq} and~\ref{eq:cov_low_freq} are computed by applying a DWT decomposition to the training data up to the appropriate wavelet level (corresponding to the level being trained) and calculating the average power spectrum or cross-spectrum of the wavelet coefficients across the entire dataset. This approach ensures that when using the CC prior, the wavelet coefficients (green boxes in Figure~\ref{fig:wavelet}) exhibit the same power spectrum as the prior or latent distribution (blue boxes in Figure~\ref{fig:wavelet}). As a corollary, we note that for the WN prior, the standard normal assumption applies in wavelet coefficient space rather than in pixel space.

Taking advantage of WF modularity, we develop a hybrid approach that applies different priors at different wavelet scales. Our Hybrid Component Correlated (HCC) prior applies a CC prior for the highest resolution wavelet scale (i.e., the finest detail coefficients \( x_{2^{n-1}}^d \sim p(x_{2^{n-1}}^d | x_{2^{n-1}})\)), while retaining the standard WN prior for coarser scales. This hybrid approach leverages the adequate performance of WN priors at large scales and eases the training of the more non-Gaussian structure at small scales with the CC prior.

\section{Implementation}
\label{sec:imple}
\subsection{Training Data and preprocessing}
Our training data come from the \texttt{Agora} full-sky extragalactic foreground simulations~\cite{yuuki} generated given the N-body dark-matter simulation MDPL2~\cite{mdpl2}.
The available foreground maps include CIB, tSZ, kSZ, and radio sources, all of which are correlated with the $\kappa$ field. 

The \texttt{Agora} simulations generate maps through a physically-motivated approach tailored to each component. For $\kappa$, the simulation employs a hybrid technique combining multi-plane ray-tracing for lower redshifts ($z \leq 8.6$) with the Born approximation for higher redshifts. Ray-tracing through the MDPL2 $N$-body simulation captures nonlinear lensing effects, while high-redshift contributions ($z > 8.6$) are modeled via Gaussian realizations of the linear power spectrum.

For the CIB component, the \texttt{Agora} simulations use the UniverseMachine catalog to assign infrared emission properties, including star formation rates (SFRs) and stellar masses, to halos, with parameters constrained by observational measurements of galaxy population statistics. The emission is integrated from $z = 0$ to $z = 8.6$, with primary contributions from dusty star-forming galaxies at $z \sim 1$--$3$. The simulation models CIB at \textit{Planck} frequencies 353, 545, and 857 GHz, calibrated against observational data from \textit{Planck}. CIB maps at 150 and 220 GHz are generated via spectral extrapolation using a modified blackbody model with a power-law transition ---meaning that above frequency $\nu'$ the SED is dominated by a power-law and below $\nu'$ it is dominated by a modified blackbody.
In this work we use the 150 GHz CIB maps for training our models.

Both $\kappa$ and CIB fields encode rich information about the universe's structure and evolution, with their properties directly shaped by the cosmological and astrophysical assumptions in the \texttt{Agora} simulations. The $\kappa$ maps are primarily determined by the cosmological parameters adopted in the simulations (specifically, those from Planck 2013). For instance, parameters like $\Omega_m$ and $\sigma_8$ influence the amplitude and scale-dependence of the matter power spectrum $C_\ell^{\phi\phi}$, which in turn sets how features such as convergence peaks and voids appear and cluster at the map level across angular scales~\cite{planks2018lensing}. The CIB maps, meanwhile, are also shaped by astrophysical inputs to the simulations. At the map level, the spatial distribution and brightness of infrared sources are established by assigning star formation rates and stellar masses to galaxies according to the UniverseMachine framework. These assignments directly control the relative brightness of individual dusty galaxies and the diffuse background~\cite{Behroozi2020}.

The input training maps are 256$\times$256 0.5-arc-minute pixels cutouts projected from the full sky maps using \texttt{pixell}.\footnote{https://github.com/simonsobs/pixell} 
The pair of inputs maps have dimensions of [2, 256, 256], corresponding to sub-regions of approximately 2$\times$2 degrees on the sky. The training dataset consists of 52,000 samples, with an additional 13,000 samples reserved for validation.

Before training, we apply several preprocessing steps to improve model performance. First, we log-transform the CIB intensity values using the equation $\text{CIB}_{\text{transformed}} = \text{sign}(\text{CIB}) \cdot \log(1 + |\text{CIB}|)$ as suggested in~\cite{mmDL}, which helps to ``Gaussianize'' the skewed CIB distribution. Both the transformed CIB and the original $\kappa$ maps are then standardized to zero mean and unit standard deviation across the entire training set. We reverse both transformations before comparing our results to the validation set. Following techniques from~\cite{waveletflowCosmology}, we also add a small Gaussian noise component to the $\kappa$ maps after the Nyquist frequency to help with training, particularly for the high-frequency components. We observed that training for the smallest scale DWT decomposition was performing poorly without this step, and adding noise significantly improves training convergence. Since these modes beyond the Nyquist frequency are not used in subsequent analysis, this preprocessing step does not impact the scientifically relevant scales.

\subsection{Model Architecture and Training}

We model the distributions of wavelet coefficients \( p(x_{2^{n-1}}^d \mid x_{2^{n-1}}) \) \dots \( p(x_{2^{1}}^d \mid x_{2^{1}}) \), \( p(x_{2^1}) \) using Normalizing Flows with a modified Glow architecture.
As discussed in Section~\ref{sec:glow}, the implementation excludes the multiscale squeeze/split operation from the original Glow architecture, instead utilizing wavelet decomposition for cross-scale representation~\cite{waveletflowCS, WaveletValiuddin2022EfficientOD}.

Our implementation trains eight probability distributions derived from a 7-level DWT decomposition: \( p(x_{2^1}) \) for the low-frequency 2×2 base map (training level 1), and the conditional distributions \( p(x_{2^1}^d \mid x_{2^1}) \), \( p(x_{2^2}^d \mid x_{2^2}) \), \( p(x_{2^3}^d \mid x_{2^3}) \), \( p(x_{2^{4}}^d \mid x_{2^{4}}) \), \( p(x_{2^{5}}^d \mid x_{2^{5}}) \), \( p(x_{2^{6}}^d \mid x_{2^{6}}) \), and \( p(x_{2^{7}}^d \mid x_{2^{7}}) \) for detail coefficients (training levels 2-8). Each training level uses a Glow model with affine coupling layers adjusted to the complexity requirements of each resolution scale: 16 layers for the four highest-resolution training levels (5-8), and 8 layers for the four lowest-resolution training levels (1-4). Each affine coupling layer contains a three-step convolutional network: a 3×3 convolution, followed by a 1×1 convolution, and another 3×3 convolution, all with a hidden dimension of 256. We implement checkerboard pattern data splitting in the affine coupling network, use invertible \(1\times1\) convolutions for permutations, and set the activation normalization scale to 1.0.

Our training strategy addresses the distinct challenges presented by the target distribution at different resolution scales. The coarsest scales (training levels 1-4) have smaller spatial dimensions and exhibit higher statistical variance. We train these levels using batch sizes of 1024 to reduce variance per batch. For the four finest resolution levels (training levels 5-8), which have progressively larger spatial dimensions, we use decreasing batch sizes of 512, 256, 128, and 64 respectively to manage GPU memory constraints. The two finest resolution levels (levels 7 and 8) additionally use single-precision floating point (float32) for memory efficiency.

All models are trained on single Tesla A100 GPUs using the Adam optimizer with a learning rate of 0.001. We implement an early stopping criterion that terminates training after ten consecutive epochs without improvement in epoch-averaged loss. Training duration varies by decomposition level: the coarsest-scale level converges within 4 hours ($\sim 450$ epochs), while the finest-scale level requires approximately 60 hours ($\sim 150$ epochs).

\section{Results}
\label{sec:Results}

Our \wf\ models were trained on $\kappa$-CIB maps using the HCC prior, which employs a WN prior for all scales except the highest frequency wavelet component where a CC prior is used. Additional information on prior selection are included in Appendix~\ref{sec:appendix}.

In Figures~\ref{fig:map_samples},~\ref{fig:power_spectra}, and~\ref{fig:minkowski_functionals}, we present a comparative analysis of the performance of our model. Figure~\ref{fig:map_samples} displays sample $\kappa$ and CIB maps from the validation set and generated from the trained model. The maps demonstrate good qualitative agreement between the validation maps and the \wf-generated samples. 
Our model effectively reproduces the spatial fluctuations present in the $\kappa$ validation maps, as well as the clustered emission patterns observed in the CIB validation maps. 
The correlation between the $\kappa$ and CIB fields are successfully learned as well, as evidenced by the higher emission in the CIB map where the convergence field is positive (corresponding to matter overdensities).

\begin{figure}[htbp]
    \centering
    \includegraphics[width=0.6\textwidth]{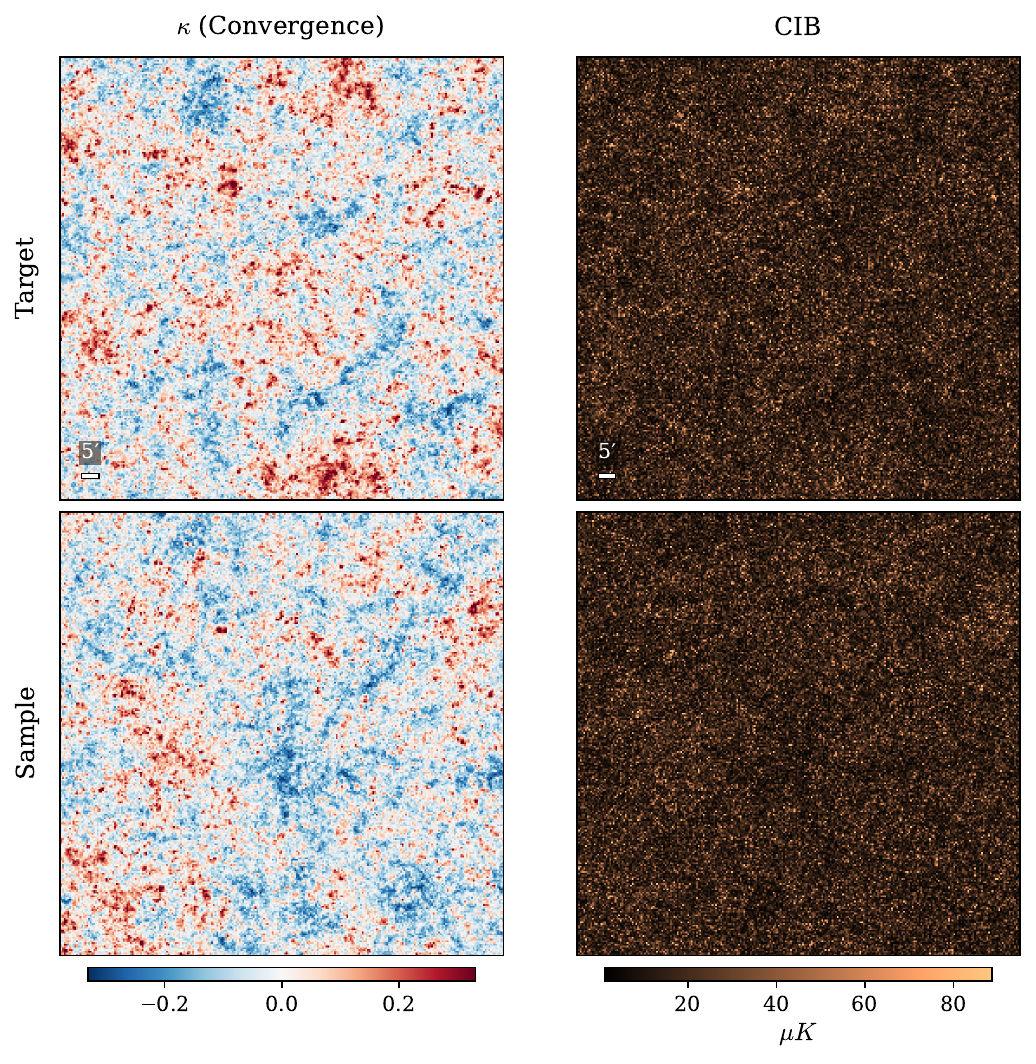}
    \caption{Example maps showing the lensing convergence field ($\kappa$, left) and cosmic infrared background (CIB, right). The top row displays \textit{Agora} simulation maps from the validation set, while the bottom row shows corresponding samples generated by our model.
    There is good qualitative agreement between the generated samples and those from the validation set.}
    \label{fig:map_samples}
\end{figure}

\begin{figure}[htbp]
    \centering
    \includegraphics[width=0.95\textwidth]{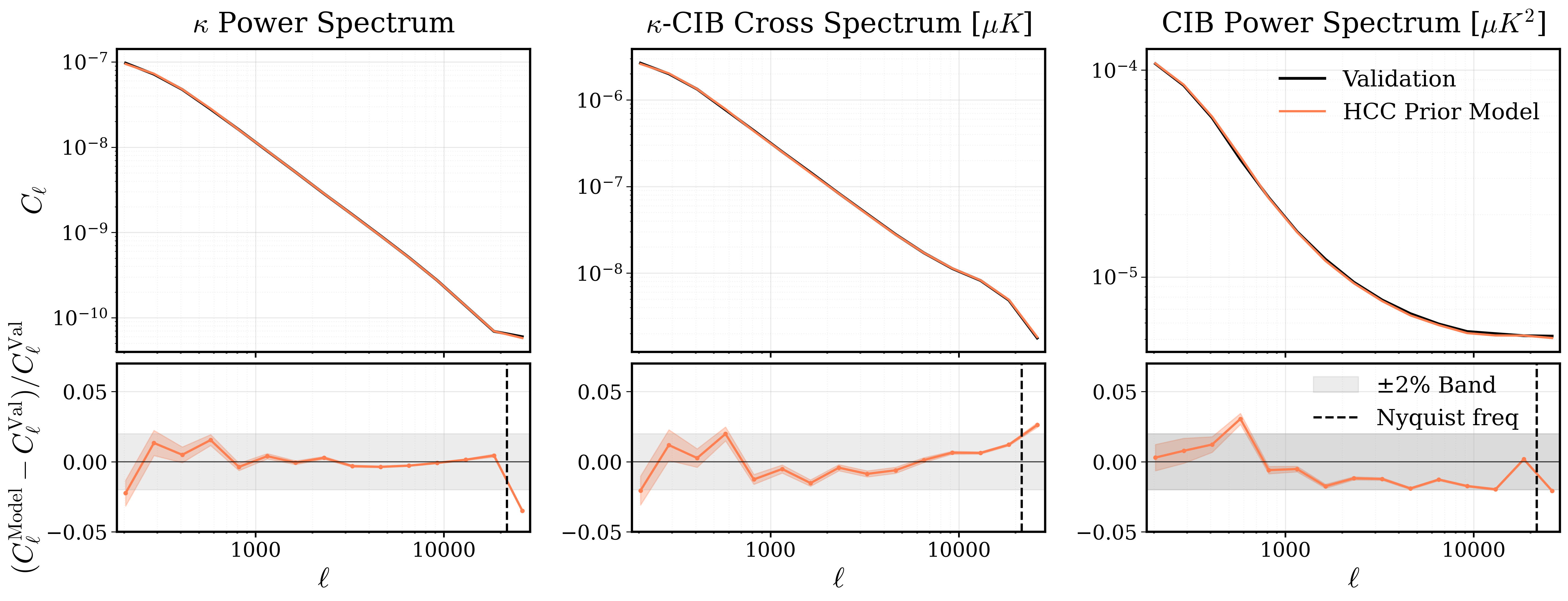}

    \caption{\small (Top Panel) Power spectra of validation maps and those generated by our fiducial model with HCC prior, showing auto-spectra for $\kappa$ and CIB along with their cross-spectrum. (Bottom Panel) Fractional difference between the mean power spectra for the validation and model generated maps. The mean power spectra are calculated by averaging over 13,000 samples for both the validation maps and the model-generated maps. The orange shaded region represents the standard (1\(\sigma\)) error on the fractional difference, propagated from the standard errors of the model-generated and validation spectra.} 
    \label{fig:power_spectra}
\end{figure}

\begin{figure}[htbp]
\centering
\begin{subfigure}{\textwidth}
\begin{center}
\includegraphics[width=0.99\textwidth]{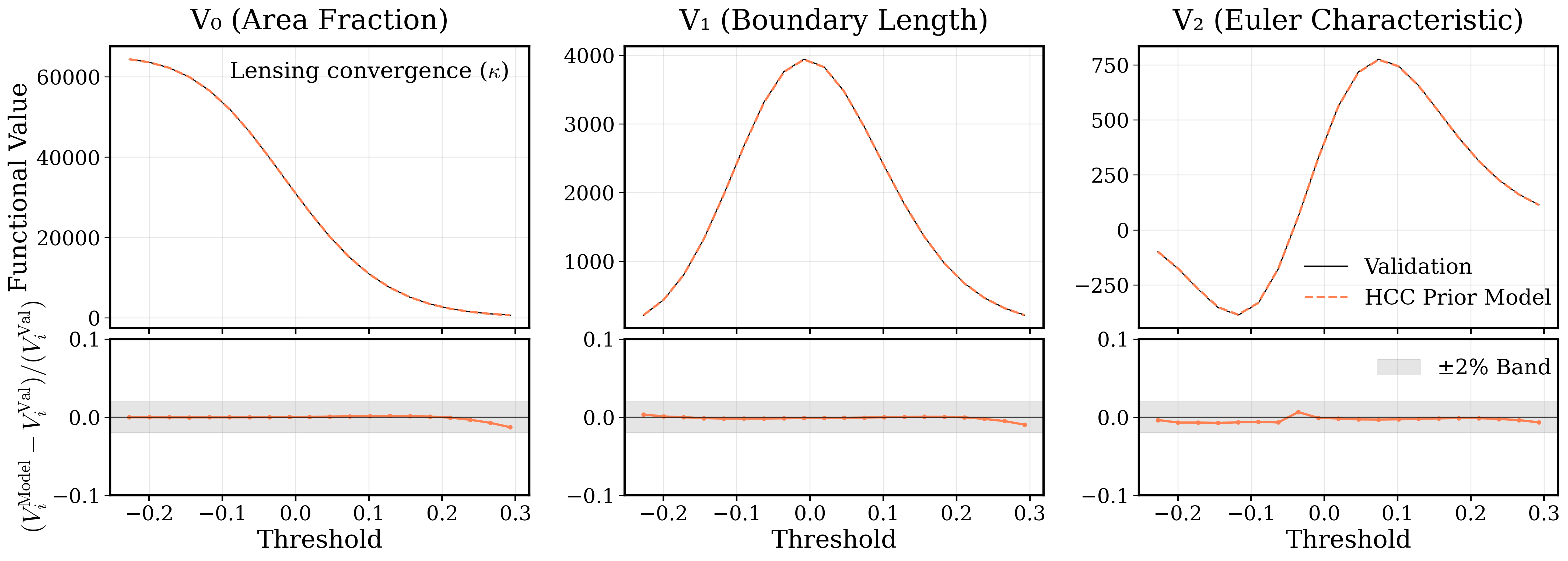}
\end{center}
\end{subfigure}

\begin{subfigure}{\textwidth}
\begin{center}
\includegraphics[width=0.99\textwidth]{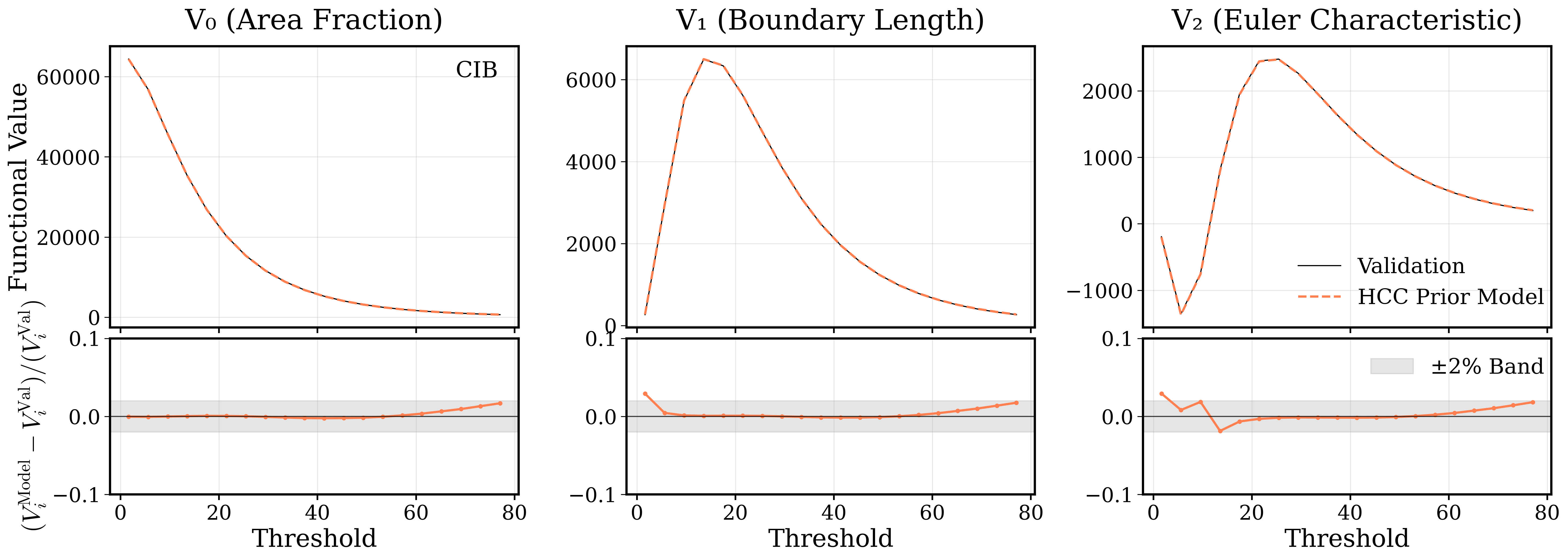}
\end{center}
\end{subfigure}

\caption{Minkowski functionals and fractional differences for $\kappa$ (Top Panel) and CIB (Bottom Panel). The solid black lines represent the validation data, while the dashed orange lines show the model predictions. The model-generated samples statistically possess non-Gaussian fluctuations described by Minkowski functionals consistent with the \texttt{Agora} simulations.}
\label{fig:minkowski_functionals}
\end{figure}
Figure~\ref{fig:power_spectra} quantifies our model's accuracy through power spectrum analysis. The results show that our model accurately reproduces the power-spectrum-level properties of \texttt{Agora} $\kappa$ and CIB maps across multiple scales. The $\kappa$ power spectrum bias remains within 1\% for most multipoles up to the Nyquist frequency, with marginal increases to 2\% at $\ell < 700$. Similarly, the $\kappa$-CIB cross-spectrum maintains a bias below 2\% across all scales up to the Nyquist frequency. The CIB power spectrum shows comparable accuracy, with a bias of up to 2\% across most multipoles, only marginally increasing to 2.5\% in the range of $\ell \approx 400-600$.

Figure~\ref{fig:minkowski_functionals} displays Minkowski functionals calculated using 20 bins between the 1st and 99th percentiles of the data. The model-generated outputs show good agreement with validation values, remaining within 1\% for all $\kappa$ functionals and within 2.5\% for all CIB functionals. This agreement confirms that our model successfully captures non-Gaussian characteristics beyond two-point statistics.

\section{Discussion and Conclusions}
\label{sec:DandC}
We have demonstrated that \wf\ models with scale-dependent priors effectively capture the joint statistics of correlated $\kappa$-CIB fields. Power spectra of generated fields maintain accuracy within 2\% across most scales, with only CIB showing slightly higher bias (2.5\%) at large scales. The model also reproduces non-Gaussian features, with Minkowski functionals matching validation samples within 2.5\% for both fields. These results establish the viability of flow-based approaches for accurately generating correlated CMB foreground and $\kappa$ fields.

Having few-percent accuracies in the model power spectrum on all scales for $\kappa$ is relevant for current and near-future CMB experiments. 
The lensing power spectrum measurements currently have uncertainties given typical bin sizes that are $\gtrsim$ 10\%  per bin~\cite{apslensing}.
Our model will contribute a small fraction of the measurement uncertainty given our accuracy. 
Additionally, our model will have differentiating power for biases from foregrounds to the lensing spectrum measurement, which are of order 5-10\%~\cite{yuuki, vanengelenfgbias}. 
To the best of our knowledge, this model has the highest accuracy among published generative models on CMB foregrounds and secondaries~\cite{mmDL, prabhu_learning_2025}.

Prior to application to real data analysis, we plan to address several limitations of the current setup. This model needs to incorporate additional foreground components, including tSZ, kSZ, and radio sources. Our 0.5-arcminute resolution is compatible with current and upcoming large-aperture CMB experiments, with the Atacama Cosmology Telescope~\cite{Act-CMB-Spectra}, South Pole Telescope~\cite{SPT-3g-2}, the Simons Observatory~\cite{SO-forcast}, and CMB-S4~\cite{CMB-S4-Planning} all having $\mathcal{O}(1^{\prime})$ resolution. However, our current 2°×2° maps are relatively small compared to the sky coverage of these experiments, which span hundreds to thousands of square degrees. Methods like tiled cutouts~\cite{mmDL}, flow-based patching procedures~\cite{Rouhiainen:2022cwd}, and iterative outpainting~\cite{Rouhiainen:2023ewv} offer potential solutions for generating larger continuous maps.
Another limitation of our current approach is training on simulations with fixed cosmological and astrophysical parameters. Incorporating simulations with varying parameters will be  important for assessing the robustness during real data applications, where the underlying cosmological/astrophysical parameters are unknown.

Additionally, we plan to explore more expressive Normalizing Flow architectures, such as spline flows~\cite{spline}, and investigate the integration of diffusion models with wavelet decomposition techniques. These developments aim to enhance model expressivity while maintaining the computational efficiency needed for practical applications in CMB data analysis pipelines.

With a simulator that generates CMB foreground fields with the expected correlation with the lensing convergence, this work represent a step towards field-level simulation-based inference with CMB data when foregrounds contribute non-negligibly to the maps. 
This can improve upon existing approaches that maximizes the lensed CMB posterior to obtain best-fit samples of unlensed CMB maps, lensing potential, and parameters of interests~\cite{millea_a_w, diffdelensing, muse3g, museGaussT}, as none of the current methods have incorporated non-Gaussian model of foregrounds. 
Looking farther, in a future where we jointly simulate galaxies and  Galactic HI emissions with our $\kappa$-CIB fields as part of an inference, analogous CIB maps to those from~\cite{chiang2023} and~\cite{plankpr4cib, plankcib2} can be obtained as part of the output. This work represents a step towards many potential applications of field-level inference of non-Gaussian, correlated fields in cosmology and astrophysics.

\newpage
\appendix
\section{Prior selection}
\label{sec:appendix}
We conducted experiments to determine the fiducial prior choice for our baseline \wf\ model presented in the main body of this paper. These experiments led to our fiducial HCC setting, which uses WN priors for the first seven training levels (corresponding to the distributions \( p(x_{2^1}) \), \( p(x_{2^1}^d \mid x_{2^1}) \), ..., \( p(x_{2^{6}}^d \mid x_{2^{6}}) \)) and CC priors exclusively for the highest frequency wavelet coefficient (corresponding to the distribution \( p(x_{2^{7}}^d \mid x_{2^{7}}) \)).

\begin{figure}[htbp]
    \centering
    \includegraphics[width=0.98\textwidth]{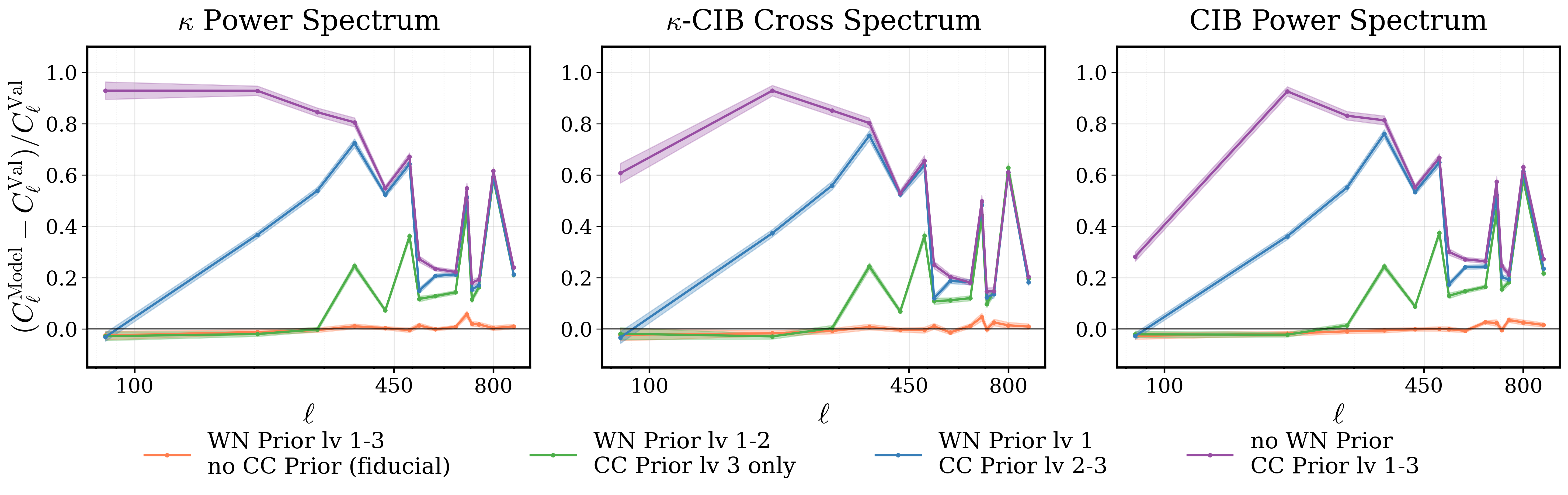}
    \caption{
        Fractional difference between the mean power spectra of validation maps and those generated by models with different prior configurations for partial map reconstructions up to the third training level (8$\times$8 maps) for $\kappa$ (left), $\kappa$-CIB (center), and CIB (right).
        Four prior configurations are compared: CC prior for training levels 1--3 (Green), WN prior for leve1 1 and CC prior for levels 2--3 (Blue), WN prior for level 1--2 and CC prior for level 3 (Purple), and WN prior for all 3 levels (Orange). 
        The shaded regions representing the $1\sigma$ standard error. The mean power spectra are calculated by averaging over 13,000 samples for both the validation maps and the model-generated maps.
        This shows that WN priors perform better at coarse scales than CC priors.
    }
    \label{fig:8x8-map}
\end{figure}

In this Section, we summarize our findings comparing different combintations of WN and CC priors across different wavelet decomposition levels. 
Figure~\ref{fig:8x8-map} shows that in our experiments, CC prior models perform worse for the coarse training levels. 
For these levels, models using CC priors exhibit power spectrum biases approaching 100\%, while models using WN priors maintain biases under 2\%. The origin of the CC prior's poor performance on large scales is not well understood, though contributing factors may include higher per-sample variance at coarse levels or nuances in the implementation. Nevertheless, this suggests that for our specific implementation and dataset, WN priors are more suitable for modeling the coarsest wavelet coefficients.

\begin{figure}[htbp]
    \centering
    \includegraphics[width=0.98\textwidth]{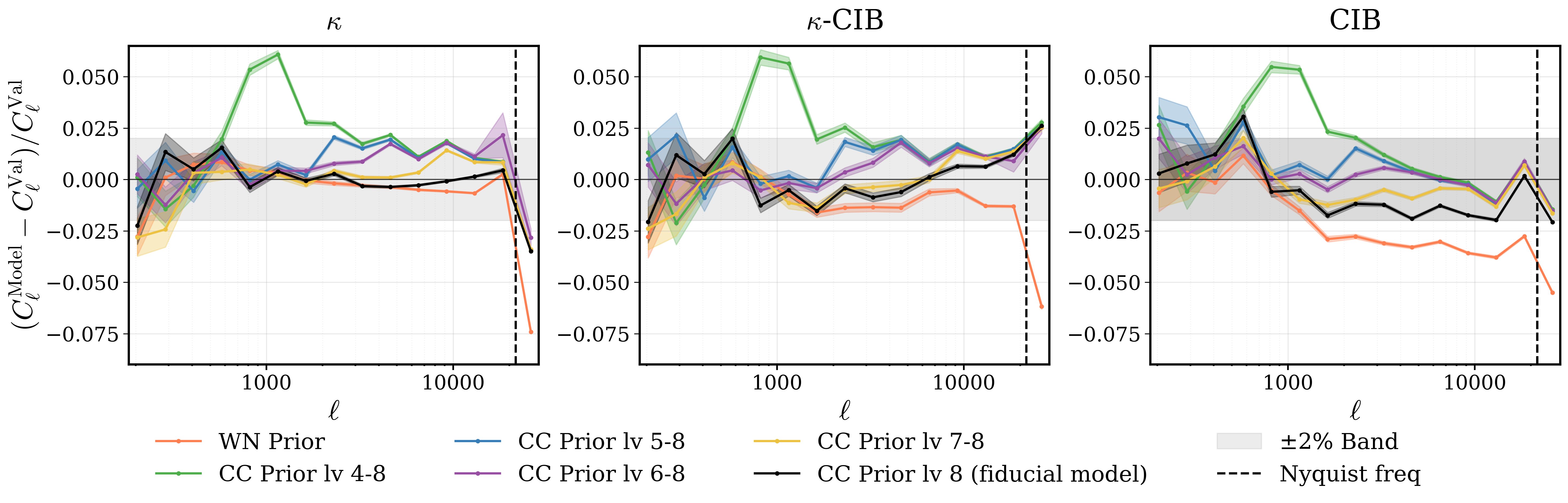}
    \caption{
        Fractional difference between the mean power spectra of validation maps and those generated by models with different prior configurations for full-resolution 256$\times$256 maps for the
        $\kappa$ (left), $\kappa$-CIB (middle), and CIB (right) cases. 
        The figure compares WN prior for all training levels (Orange) against CC priors applied to varying numbers of high-frequency levels: level 8 only (Black; fiducial), levels 7--8 (Yellow), levels 6--8 (Purple), levels 5--8 (Blue), and levels 4--8 (Green). 
        All remaining training levels use WN priors. 
        The optimal configuration uses CC priors at least for level 8, with the option of extending up to level 5, maintaining biases below 2.5\% across all scales and components. 
        The shaded regions represent the $1\sigma$ standard error. The mean power spectra are calculated by averaging over 13,000 samples for both the validation maps and the model-generated maps. The vertical dashed line indicates the Nyquist frequency.
    }
    \label{fig:full-256x256-map}
\end{figure}

\begin{figure}[htbp]
\centering
\begin{subfigure}{\textwidth}
\begin{center}
\includegraphics[width=0.95\textwidth]{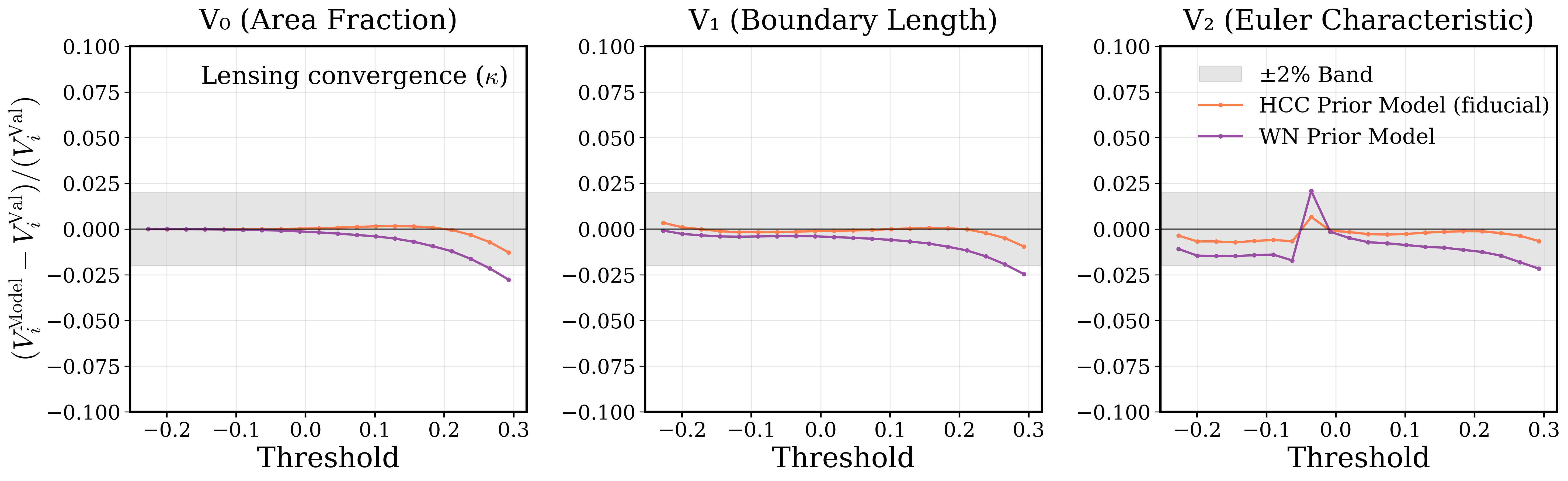}
\end{center}
\end{subfigure}

\begin{subfigure}{\textwidth}
\begin{center}
\includegraphics[width=0.95\textwidth]{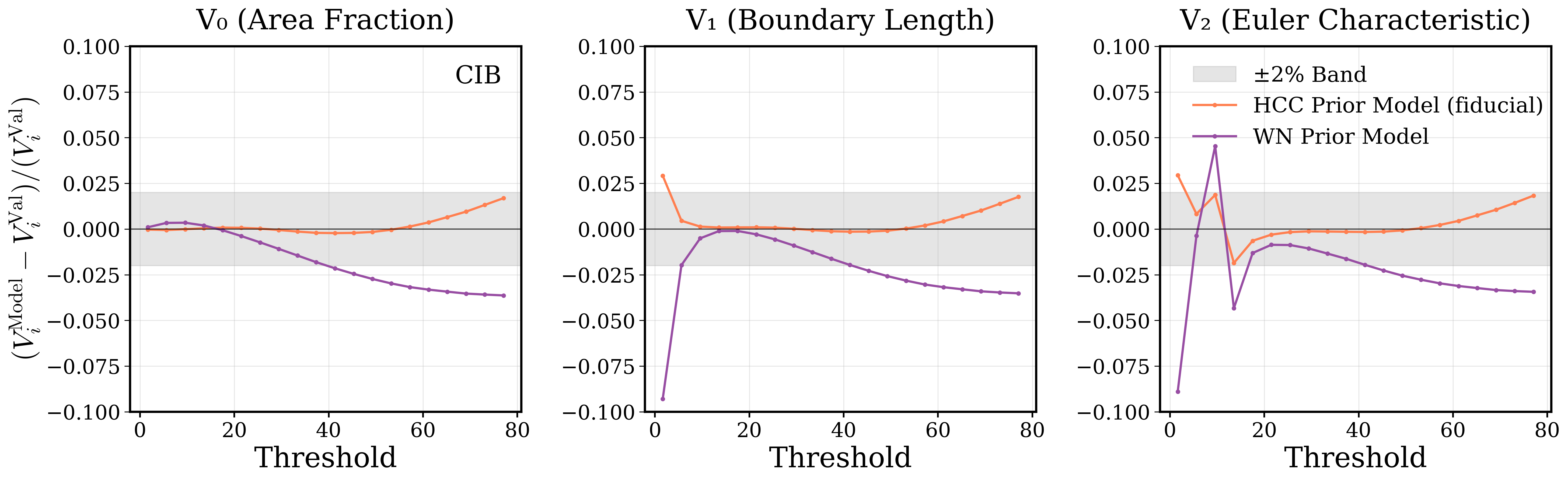}
\end{center}
\end{subfigure}

\caption{
Fractional difference of the Minkowski functionals for $\kappa$ (Top Panel) and CIB (Bottom Panel). 
The dashed orange line corresponds to the fiducial HCC prior model and dashed purple line represents the WN prior model.
This shows that the HCC prior better captures non-Gaussian statistics in the maps compared to the WN prior.}
    \label{fig:minkowski_functionals_wnvshcc}
\end{figure}

In contrast, Figure~\ref{fig:full-256x256-map} reveals the benefits of CC priors at the finest level of DWT decomposition. 
When applied specifically to the highest frequency coefficients (i.e. \(x_{2^{7}}^d \sim  p(x_{2^{7}}^d \mid x_{2^{7}}) \)) while using WN priors elsewhere, the models achieved an improvement of up to 3\% in power spectrum recovery at small scales for both $\kappa$ and CIB.
The intermediate decomposition levels (i.e. \(x_{2^{4}}^d \sim p(x_{2^{4}}^d \mid x_{2^{4}}) \) through \(x_{2^{6}}^d \sim p(x_{2^{6}}^d \mid x_{2^{6}}) \)) showed negligible differences between the two prior types. Figure~\ref{fig:minkowski_functionals_wnvshcc} further demonstrates that the HCC approach better captures non-Gaussian features, with the maximum bias in Minkowski functionals decreasing from approximately 10\% to 2.5\% compared to WN-only models. Based on these findings, we adopted the HCC prior, a scale-dependent prior strategy that takes advantage of the strengths of each prior type.

\clearpage
\acknowledgments
The authors would like to thank Sean Gasiorowski for useful discussion and comments on the manuscript, and Giuseppe Puglisi and Moritz Muenchmeyer for feedback on this work.

Our work builds upon the PyTorch implementation of the \wf model from Ref~\cite{WaveletValiuddin2022EfficientOD}. We also acknowledge the following publicly available Python packages used in this work: \textit{matplotlib}~\cite{Matplotlib}, \textit{numpy}~\cite{2020NumPy-Array}, \textit{pandas}~\cite{pandas}, \textit{quantimpy}~\cite{quantimpy}, \textit{scikit-image} ~\cite{scikit-image}, \textit{scipy}~\cite{2020SciPy-NMeth}, \textit{seaborn}~\cite{seaborn}, \textit{py-lmdb}~\cite{lmdb_python}, \textit{pytorch} ~\cite{pytorch}, \textit{pytorch-wavelets}~\cite{pytorch_wavelets}, and \textit{PyWavelets}~\cite{Pywavelets}.

This work used the resources of the SLAC Shared Science Data Facility (S3DF) at SLAC National Accelerator Laboratory. S3DF is a shared High-Performance Computing facility, operated by SLAC, that supports the scientific and data-intensive computing needs of all experimental facilities and programs of the SLAC National Accelerator Laboratory. SLAC is operated by Stanford University for the U.S. Department of Energy's Office of Science.
This research used resources of the National Energy Research Scientific Computing Center (NERSC), a Department of Energy Office of Science User Facility.

MM and WLKW acknowledge support from an Early Career Research Award of the Department of Energy and a Laboratory Directed Research and Development program as part of the Panofsky Fellowship program at the SLAC National Accelerator Laboratory. 
The SLAC authors acknowledge support by the Department of Energy, under contract DE-AC02-76SF00515.


\bibliographystyle{JHEP}
\bibliography{biblio}

\end{document}